\documentclass[conference]{IEEEtran}
\ifCLASSOPTIONcompsoc
  \usepackage[nocompress]{cite}
\else
  \usepackage{cite}
\fi
\usepackage{comment}
\usepackage{url}
\usepackage{graphicx}
\usepackage[table]{xcolor}
\usepackage{pifont}
\usepackage{algorithm}
\usepackage{algpseudocode}
\usepackage{amsmath}
\usepackage{comment}
\usepackage{multirow}
\usepackage{graphicx}
\usepackage{booktabs}
\usepackage[table]{xcolor}
\usepackage{array}

\usepackage{cite}                     
\usepackage{url}

\ifCLASSOPTIONcompsoc
  \usepackage[caption=false,font=footnotesize,labelfont=sf,textfont=sf]{subfig}
\else
  \usepackage[caption=false,font=footnotesize]{subfig}
\fi

\newcommand{\authnote}[2]{{\bf[#1: #2]}}

\renewcommand{\paragraph}[1]{\vspace{0.1in}\noindent{\bf{#1}.}}

\ifCLASSINFOpdf
\else
\fi
\begin{document}
%
\title{A Non-Line-of-Sight, Multi-Modality-based Side-Channel IP Theft Attack on Additive Manufacturing Using Dual Smartphones}

\author{
\IEEEauthorblockN{Amirhossein Jamarani\IEEEauthorrefmark{1}, Diba Afroze\IEEEauthorrefmark{1}, Yazhou Tu\IEEEauthorrefmark{2}, Mark Yampolskiy\IEEEauthorrefmark{2}, Xiali Hei\IEEEauthorrefmark{1}}
\IEEEauthorblockA{\IEEEauthorrefmark{1}University of Louisiana at Lafayette, Lafayette, LA, USA\\
Email: \{amirhossein.jamarani1, diba.afroze1, xiali.hei\}@louisiana.edu}
\IEEEauthorblockA{\IEEEauthorrefmark{2}Auburn University, Auburn, AL, USA\\
Email: \{yzt0065, mzy0033\}@auburn.edu}
}


%


\renewcommand{\paragraph}[1]{\vspace{0.02in}\noindent{\bf{#1}.}}

\newcommand{\xh}[1]{{\authnote{\textcolor{blue}{Sharon}}{\textcolor{cyan}{#1}}}} 

\newcommand{\DA}[1]{{\authnote{\textcolor{violet}{Diba}}{\textcolor{violet}{#1}}}} 
\maketitle

\begin{abstract}
Additive Manufacturing (AM) has revolutionized major sectors, including aerospace, automotive, and healthcare, by enabling adjustable production. As the usage of AM increases, so does the risk of Intellectual Property (IP) leakage during the printing process due to unintended side-channel emissions.
Current studies and attack scenarios on 3D printers face three challenges: low success and accuracy rates in final G-code reconstruction, limited distance range for attacking the 3D printer's IP, and reliance on specialized, overt data-collection tools. 
This paper presents a side-channel attack that addresses the noted limitations by using two smartphones’ internal sensors. We position the smartphones 60 cm away in a non-line-of-sight setup to collect the 3D printer’s acoustic and magnetic emissions. Our attack successfully reconstructs the G-code commands of the final objects at the rate of 98.89\% on command-level reconstruction accuracy. Additionally, we evaluate the transferability of our attack strategy by applying it to another 3D printer in a different environment. Our proven unauthorized access to the reconstructed G-code and thus to the IP of the AM system indicates the security weaknesses in 3D printing, highlighting the need for mitigating side-channel attacks.
\end{abstract}

\section{Introduction}
Additive Manufacturing (AM), commonly referred to as 3D printing, has reshaped how products are designed, prototyped, and manufactured \cite{ramesh2024additive,kaur2025artificial}. By enabling rapid prototyping, on-demand production, and the fabrication of complex geometries that are difficult to achieve with conventional subtractive processes, AM has become an important driver of innovation across multiple sectors \cite{alarifi2024revolutionising}. In parallel, advances in AM education and design-for-AM tools have led to the transition of 3D printing from primarily visualization and prototyping to industrial-grade production workflows \cite{ukwaththa2024review,vinod2024advancing}. This transition is frequently associated with shorter development cycles, reduced material waste, lower production costs for suitable part classes, and increased opportunities for personalization \cite{kahhal2024recent,al2024additive,ajamloo2024review}. Consequently, 3D printing has been adopted in domains such as aerospace \cite{masrafee2025innovations}, healthcare \cite{minghetti2025regulatory}, automotive \cite{szalai2025investigation}, construction \cite{hassan2024towards}, and food printing \cite{thorakkattu20253d}. The accessibility of modern printers, combined with widely available documentation and increasingly user-friendly software stacks, has further supported this adoption \cite{kantaros20223d,rojek2024use, 10.1007/978-3-031-93354-7_3}.

\begin{figure}[ht]
    \centering
    \includegraphics[width=1.0\linewidth]{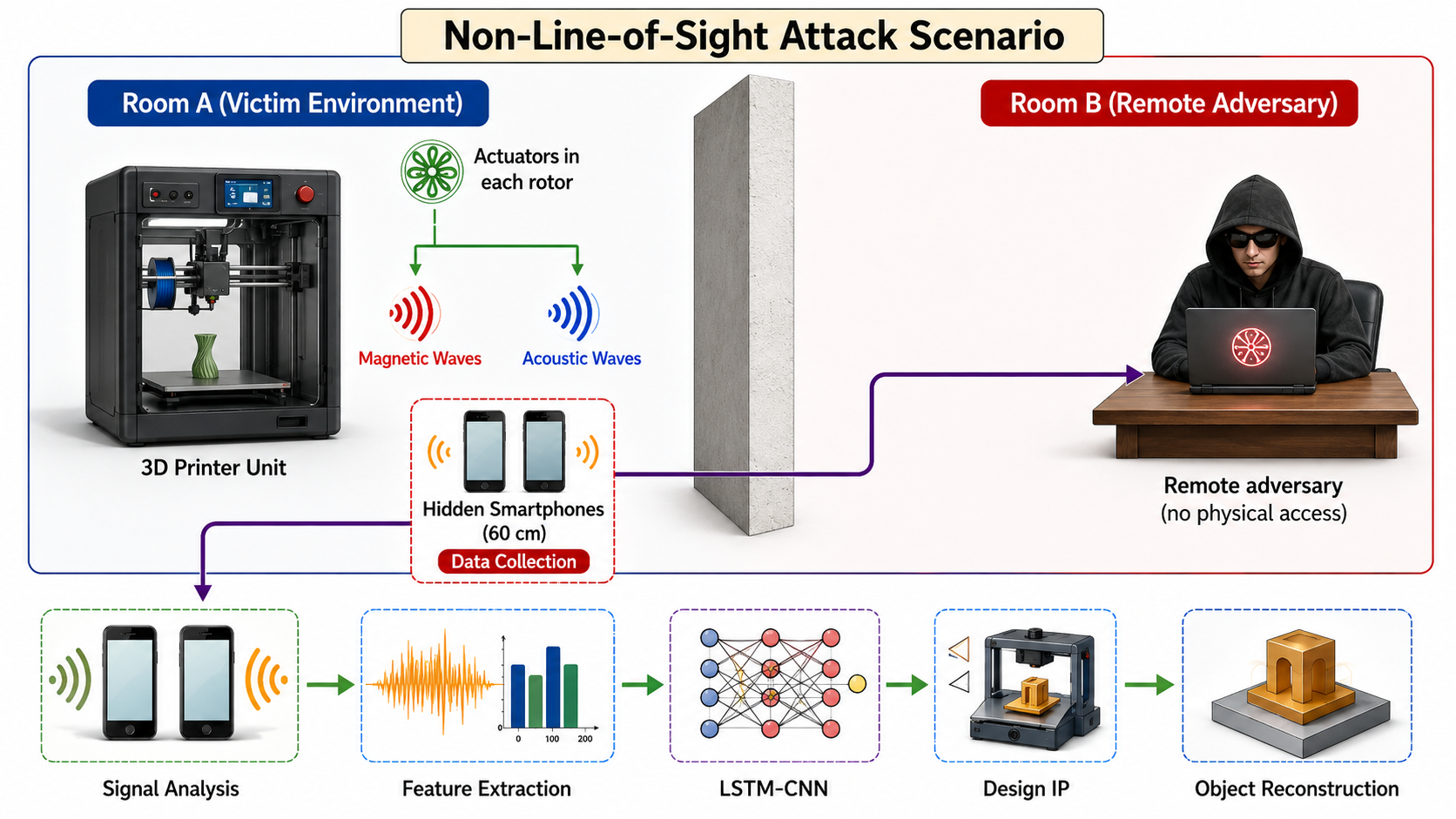}  
    \caption{End-to-end attack workflow showing how acoustic and magnetic emissions from a 3D printer are captured, processed, and used to reconstruct the underlying G-code in our study.}
    \label{fig:attack_overflow}
\end{figure}

Despite these benefits, the 3D printing process is vulnerable to side-channel attacks. Prior studies \cite{asgar2026quietprint, madamopoulos20243d,dolgavin2024stop,chen2024comprehensive,song2016my,gatlin2021encryption} have shown that G-code instructions can be partially reconstructed by exploiting unintended emissions, including acoustic and magnetic signals produced during printer operation. However, many existing evaluations rely on assumptions that can be difficult to justify outside controlled settings. These assumptions often include: (i) close-range smartphone placement within direct line-of-sight of less than 30 cm or usage of specialized data-collection equipment \cite{gatlin2021encryption,asgar2026quietprint,liang2022hiding,imran2025ensemble} (ii) simplified operational behavior in which the printer is assumed to execute only printing motions, while non-print states, for example, cooling cycles, idle periods, calibration routines, are ignored; and (iii) data collection under laboratory conditions with limited ambient noise to maximize signal fidelity. In realistic manufacturing environments, these conditions are frequently violated due to background noise, occlusion, shared workspaces, and concurrent machine activity.

In this work, we study a practical attacker model in which commodity sensing devices are used in a non-line-of-sight setting, as depicted in Figure 1. Specifically, we conceal two smartphones and leverage their built-in microphones and magnetometers to capture both acoustic and magnetic emissions from a 3D printer, with the following model specification \cite{sansing2014taz}. Our side-channel attack extends the effective range to up to 60 cm and achieves a success rate of 98.89\% on the command-level reconstruction accuracy under more realistic conditions that include background noise and a second 3D printer \cite{Creality} operating simultaneously in the same room. We also conduct a second attack scenario using a different 3D printer unit of the same brand in a different physical environment. In this scenario, the smartphones are again placed in a non-line-of-sight configuration, but at increased distances of 65 cm and 70 cm from the printer’s surface bed. The attack successfully reconstructs multiple objects with the following accuracy of the G-code commands, including a gear icon at 98.05\%, turbine blades at 98.10\%, and a wrench at 98.35\%. Finally, we discuss feasible countermeasures to reduce side-channel leakage and mitigate the risk of unauthorized recovery of sensitive manufacturing instructions. 
Our main contributions are summarized as follows:
\begin{itemize}
    \item \textbf{Dual-smartphone, multi-modal attack under realistic noise:} We develop and evaluate a concealed attack using two smartphones and a dynamic noise modeling framework that incorporates environmental noise. The approach enables G-code reconstruction in a noisy environment, including scenarios where another 3D printer operates concurrently.

    \item \textbf{Non-line-of-sight reconstruction without physical contact:} We demonstrate successful side-channel attacks with no physical contact to the printer, using non-line-of-sight phone placements at 60~cm away from the printer's surface plate.

    \item \textbf{High-fidelity reconstruction outcome:} We reconstruct G-code commands that yield a printed object matching the original, achieving 98.89\% accuracy. Importantly, unlike prior works, our method sustains high accuracy at larger attack distances, demonstrating improved robustness while also reducing dimensional and surface-quality errors.

    \item \textbf{Robustness across different attack conditions:} We conduct an additional attack in a different environment and on a different printer unit to assess whether the methodology remains effective under changed conditions and varying background noise levels.

    \item \textbf{Mitigation guidance:} We outline practical countermeasures for protecting AM processes by analyzing defenses that reduce side-channel exposure and hinder reconstruction attempts.
\end{itemize}

\section{Background}

\textbf{Additive Manufacturing.} AM is widely used to fabricate parts directly from digital designs, enabling rapid iteration and on-demand production across many domains \cite{siddiqui2025condition}. In a typical fused-filament fabrication pipeline, a designer produces a CAD model \cite{hunde2022future} and exports it as a mesh known as STL\cite{perez2025protocol}. A slicer/CAM tool \cite{kyratsis2025review} then converts this geometry into G-code \cite{lopes2026assessment}, which specifies low-level instructions such as toolhead motion, extrusion behavior, and temperature setpoints. The printer executes these commands to coordinate motion and process control during fabrication, making the G-code stream the primary representation of both design intent and manufacturing parameters.

In a typical AM execution stack, there are three main functional components \cite{hamza2025revolutionizing}. First, the printer controller, which is known as firmware, interprets G-code and computes control actions required to realize commanded trajectories and process settings. Second, sensing and feedback mechanisms, such as temperature sensing and basic state monitoring, provide measurements used to maintain stable operation. Third, actuation and power subsystems, including stepper motors, motor drivers, heaters, and extruder mechanisms, physically implement the commanded motion and thermal behavior over time. Together, these components form a cyber-physical loop that translates a digital instruction stream into a physical object.

\textbf{Side-Channel Emissions in AM.} While AM offers strong functional advantages, the same execution stack can unintentionally leak information during printing. Actuation and control operations produce observable physical emanation that correlates with the underlying command stream. In particular, stepper-motor motion generates distinctive acoustic patterns, and motor drivers and associated electronics produce magnetic emissions. These signals are not designed for external observation; however, they can be captured and analyzed to infer properties of the printer’s motion and extrusion behavior, and potentially recover command-level information.

Prior work \cite{gatlin2021encryption,al2016acoustic, di2025enhancing} has demonstrated the feasibility of reconstructing aspects of 3D printing instructions from acoustic and magnetic channels under controlled conditions, often using favorable sensor placement and low-interference environments. However, practical settings introduce background noise, occlusion, concurrent machine activity, and constrained sensor placement, all of which can degrade signal quality and limit attack viability. These constraints motivate attacker models that rely on commodity sensing devices and multi-modal capture to improve robustness under realistic conditions. In this work, we therefore consider a non-line-of-sight attacker using two smartphones to record acoustic and magnetic emissions and reconstruct G-code at the command level.  

\section{Threat Model}
\textbf{Adversary Assumptions.} The attacker aims to recover the G-code command sequence executed by a target 3D printer using only side-channel emissions. We assume the attacker has no access to the printer firmware, internal telemetry, CAD model, STL file, or original G-code of the targeted unit. The attacker may, however, possess general knowledge of the printer model and the printing process. The attacker deploys two hidden commodity smartphones equipped with microphones and magnetometers in a non-line-of-sight setting 60 cm from the printer's surface plate at a 45$^\circ$ angle to the unit. These devices are used only to record acoustic and magnetic emissions produced during printing. The attacker does not physically modify the printer, connect to its interfaces, or access its internal logs.

For the sake of evaluation only, the original G-code is used as ground truth to compare against the reconstructed command sequence. This ground truth is not available to the attacker during the attack phase.  

\textbf{Non-Invasive Side-Channel Observation.} We consider a non-invasive, touch-free, and hidden adversary as shown in Figure \ref{fig:attack_overflow}. The attacker does not modify the printer’s hardware or firmware and does not interact with any digital interfaces connected to the printer. In addition, the attacker has no access to internal telemetry such as motor control signals, sensor readings, system clocks, or device logs. The IP associated with the printing task, including the CAD model, STL file, and original G-code, is also assumed to be unavailable to the adversary. 

Instead, the attacker relies exclusively on external side-channel signals that naturally arise during the printer’s operation. These signals include the acoustic emissions generated by mechanical motion and the magnetic field fluctuations produced by the printer’s electrical components. By monitoring these signals, the attacker attempts to infer the sequence of commands executed during the printing process. Only for the purpose of attack accuracy, at the end of G-code reconstruction, we compare the line-by-line G-code commands to evaluate attack accuracy. 

\textbf{Adversary Capabilities.
}
The attacker can identify the target printer from a shared environment, for example, a public library, university lab, or makerspace. The attacker can easily purchase the same model and collect acoustic and magnetic signals as training data under their own controlled setup. Then the attacker can train a model to associate these side-channel patterns with corresponding G-code commands and reconstruct it.

The attacker is able to deploy two commodity smartphones equipped with built-in microphones and magnetometers 60 cm away from the 3D printer's surface plate in a hidden position.
These smartphones are utilized to capture both acoustic signals and magnetic field measurements produced during printer operation. 

The attacker is assumed to possess general knowledge about the operation of Fused Deposition Modeling (FDM) printers, specifically the model \cite{sansing2014taz}, including the fact that printer movements and actuator activity produce characteristic acoustic and magnetic patterns. Any knowledge about signal characteristics is assumed to be obtained from publicly available information or from experiments with similar devices.

To extract useful information from the recorded measurements, the attacker is capable of applying standard signal processing techniques such as filtering, synchronization, and noise reduction. These techniques allow the attacker to isolate signals associated with the target printer even in environments where other background noise sources may be present.

\section{Attack Methodology}

This section introduces the proposed methodology and describes the mechanisms taken to implement the side channel attack on the selected 3D printer. 

\subsection{Attack Overview}

Our methodology follows an end-to-end side-channel reconstruction pipeline composed of five main stages: (i) acoustic and magnetic data signal acquisition via non-line-of-sight smartphones, (ii) pre-processing and synchronization of the collected data out of two smartphones, corroborated with feature extraction, (iii) training with a hybrid LSTM-CNN model, (iv) unauthorized access to design IP, and (v) G-code reconstruction. 

\begin{figure}[ht]
   \centering
    \includegraphics[width=0.9\linewidth]{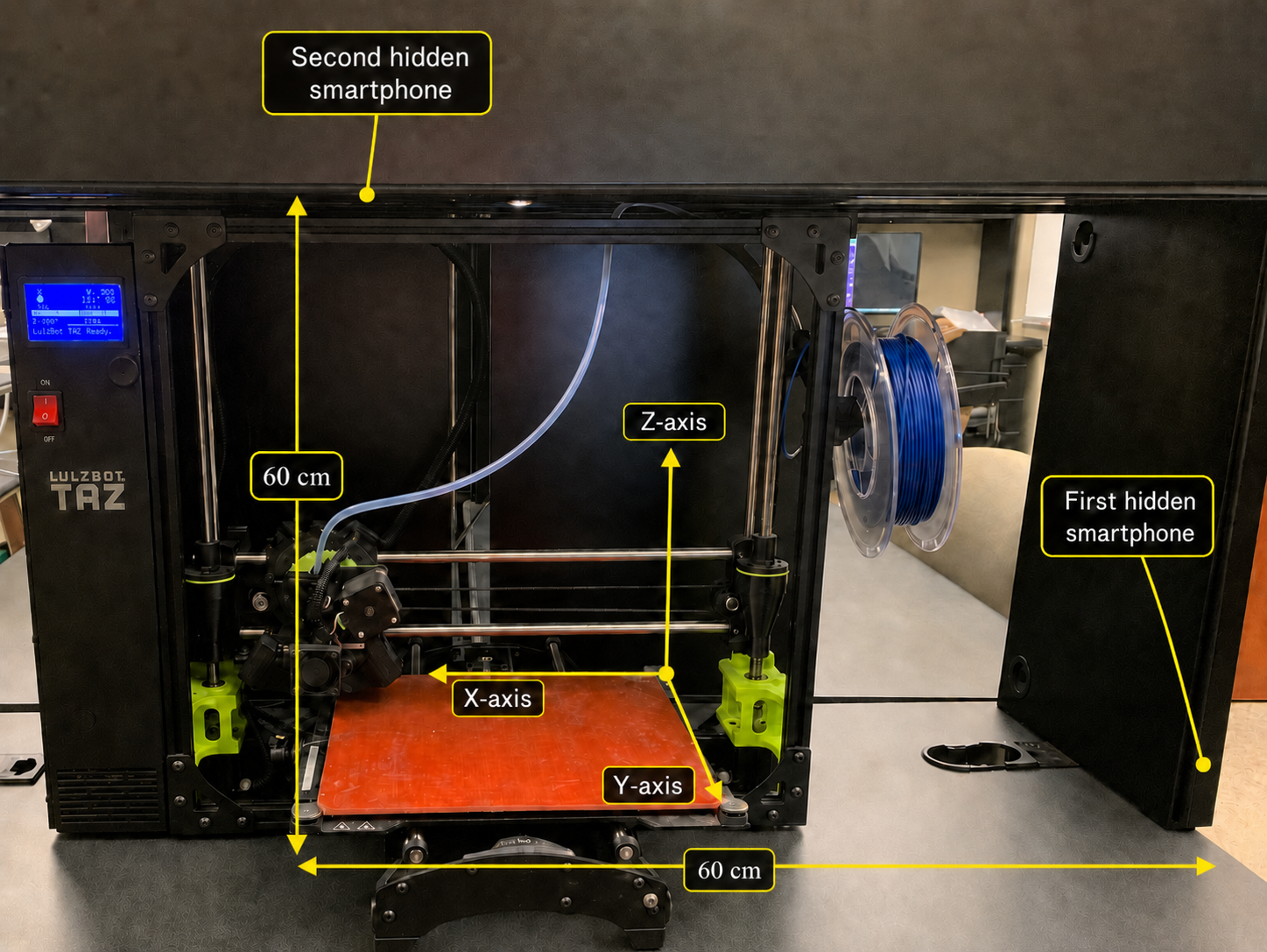}  
    \caption{Experimental setup; the smartphones are hiddenly positioned on the side and top of the table where the 3D printer is located.}
    \label{fig:Exp_setup}
\end{figure}

First, two smartphones positioned in a non-line-of-sight arrangement 60 cm away from the surface of the printer at 45$^{\circ}$ angles relative to the unit to record the printer’s unintended emissions while the victim object is being fabricated, as depicted in Figure \ref{fig:Exp_setup}.

One sensing stream captures acoustic leakage produced by mechanical motion and extrusion activity, while the second stream captures magnetic field fluctuations associated with motor drivers and moving components. Because each modality reflects a different aspect of printer behavior, combining them provides a richer representation of the ongoing print process than relying on a single channel alone.

After data collection, the recorded signals are cleaned, aligned, and segmented into analysis windows that correspond to short intervals of printer activity. From these windows, we derive features that preserve temporal variations, motion-related trends, and modality-specific patterns. The extracted features are then provided to a learning-based reconstruction module that estimates the underlying printer actions and maps them to their corresponding G-code command lines. Finally, the reconstructed commands are assembled in sequence to recover the print logic of the target object.

\subsection{Dual-smartphone Signal Acquisition}

As in equation (1), we let $S_1$ and $S_2$ denote the two smartphones used in the attack. Each phone records its own timestamped sensing stream during the printing process. For the smartphone $S_i$, the magnetometer produces a three-axis magnetic measurement, and the microphone captures an acoustic signal over time. Accordingly, the sensing streams can be represented as

\begin{equation}
\begin{aligned}
\mathbf{m}^{(i)}(t) &= \left[x^{(i)}(t),\, y^{(i)}(t),\, z^{(i)}(t)\right], \\
&\qquad a^{(i)}(t), \quad i \in \{S1,S2\}
\end{aligned}
\end{equation}

where $\mathbf{m}^{(i)}(t)$ denotes the magnetic field readings and $a^{(i)}(t)$ denotes the acoustic signal collected by the $i$-th smartphone.

In our dataset, each record contains two timestamps, one from each smartphone, together with the corresponding magnetic and acoustic measurements. This is necessary because the two devices operate independently and do not produce the same synchronized samples. As a result, the raw dataset reflects two parallel observation streams rather than a single unified timeline.

However, using two smartphones improves the practicality of the attack in two ways. First, it increases spatial coverage in a non-line-of-sight environment where one device alone cannot capture all relevant signal variations. Second, it provides observations that improve robustness when one channel is weakened by noise, device orientation, or interference from the surrounding environment. 

The 45$^\circ$ placement was also selected as a practical compromise between signal strength and directional diversity. For the acoustic channel, a simplified directional projection model suggests that the received amplitude decreases with the incident angle according to
\begin{equation}
A_r(\theta) = A_0 \cos\theta,
\end{equation}
where $A_0$ is the maximum amplitude under frontal alignment. At $\theta = 45^\circ$, the received amplitude remains
\begin{equation}
A_r(45^\circ) = \frac{A_0}{\sqrt{2}},
\end{equation}
which preserves a substantial portion of the signal while avoiding a purely frontal sensing geometry.

\begin{figure}[ht]
    \centering
    \includegraphics[width=1.0\linewidth]{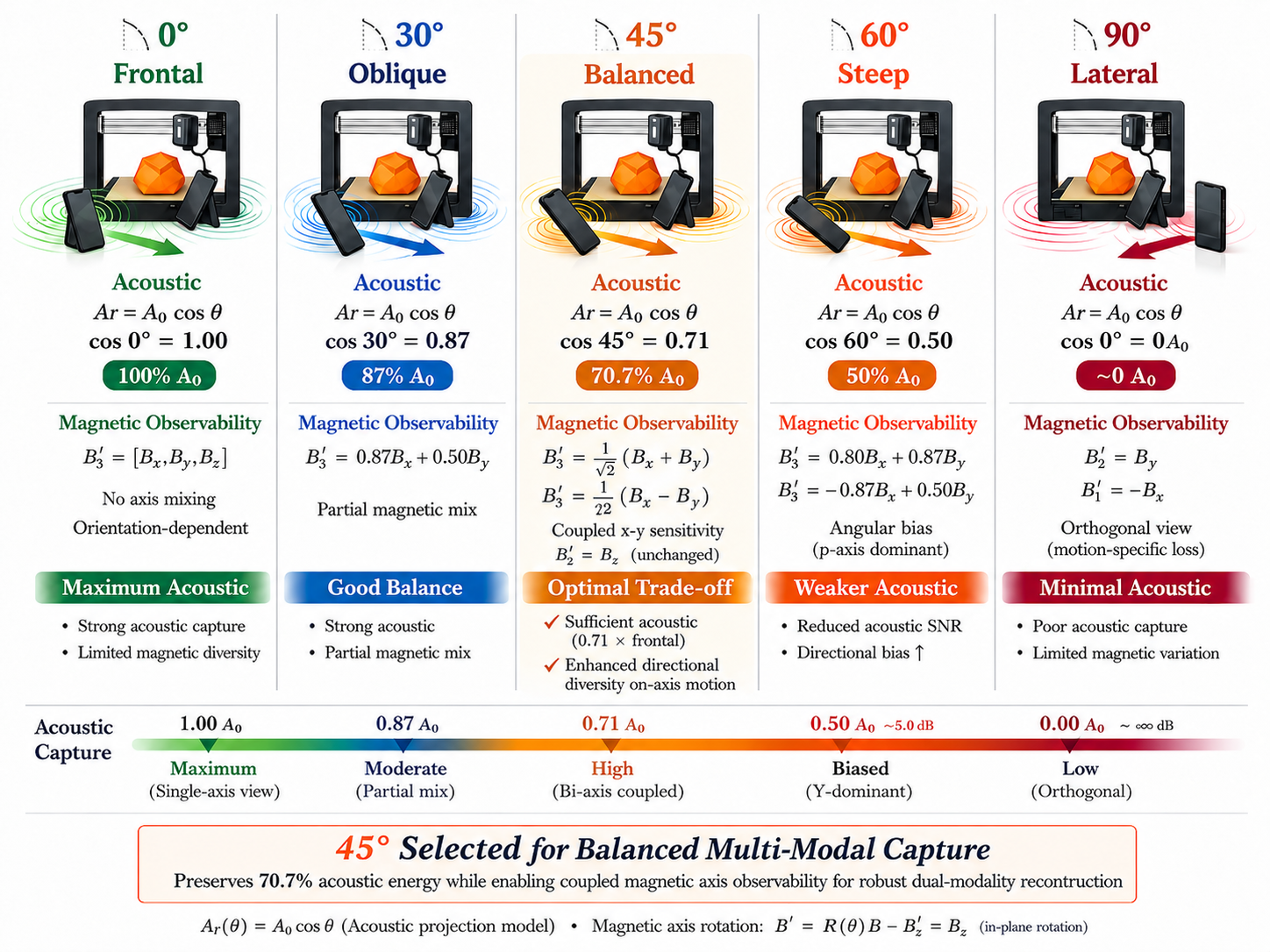}  
    \caption{Angular comparison of dual-smartphone sensing configurations for acoustic and magnetic side-channel acquisition. Acoustic signal strength follows a cosine projection model, decreasing with increasing angle, while magnetic observability improves through axis mixing under rotation. The 45° placement offers a balanced configuration, maintaining sufficient acoustic amplitude and enhanced directional sensitivity in magnetic measurements, supporting effective multi-modal G-code reconstruction.}
    \label{fig:Angular_comparison}
\end{figure}

For the magnetic channel on the inner sensing tools of the smartphones, rotating the sensing frame by an angle $\theta$ produces
\begin{equation}
B'_x = B_x \cos\theta + B_y \sin\theta,
\end{equation}
\begin{equation}
B'_y = -B_x \sin\theta + B_y \cos\theta,
\end{equation}
\begin{equation}
B'_z = B_z.
\end{equation}
At $\theta = 45^\circ$, the transformed components become
\begin{equation}
B'_x = \frac{B_x + B_y}{\sqrt{2}}, \qquad
B'_y = \frac{B_y - B_x}{\sqrt{2}}, \qquad
B'_z = B_z.
\end{equation}
This configuration captures coupled variations from both horizontal axes rather than emphasizing only a single direction as smartphones' inner magnetic sensors are located on the sides of the units. Accordingly, the 45$^\circ$ setup was used to provide a balanced multi-modal sensing geometry that preserves informative acoustic measurements while improving magnetic observability across printer motion directions, as shown in Figure \ref{fig:Angular_comparison}. These analytical expressions are intended to provide intuition regarding sensor placement and signal behavior. In practice, acoustic propagation and magnetic field measurements are influenced by environmental factors such as reflections, occlusions, and device-specific characteristics. Therefore, the effectiveness of the chosen configuration is validated empirically in our experiments.

\subsection{Signal Preprocessing and Synchronization}

Since the sensing streams are recorded independently, the first step is to align them onto a common timeline. This synchronization ensures that the magnetic and acoustic measurements from both smartphones describe the same printing interval. After alignment, the signals are normalized and divided into fixed-length windows for downstream learning.

\begin{figure}[ht]
    \centering
    \includegraphics[width=1.0\linewidth]{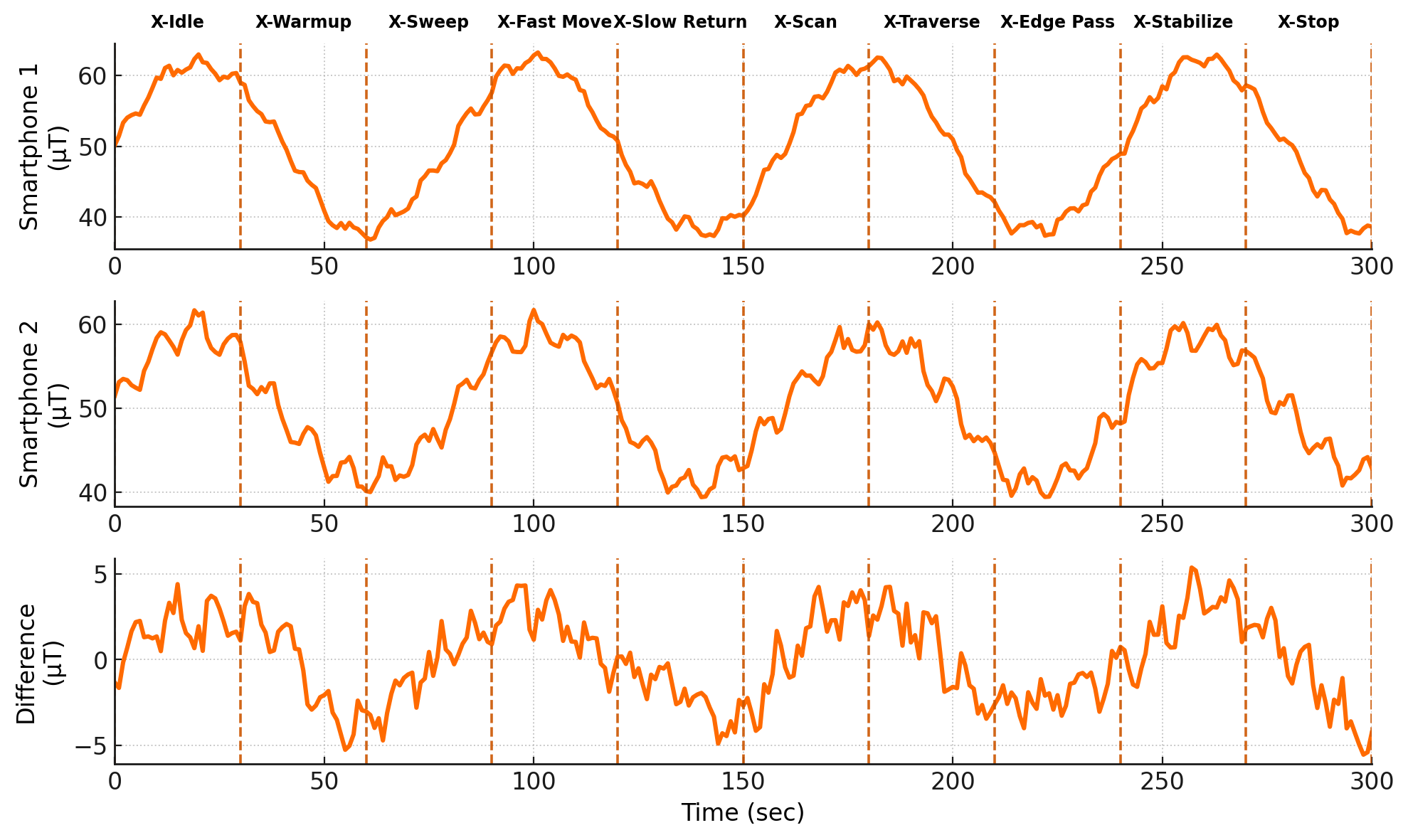}  
    \caption{Magnetic field measurements recorded by two smartphones during isolated X-axis motion. The upper and middle panels show the raw magnetic signatures captured by Smartphone 1 and Smartphone 2, respectively, while the lower panel presents their main difference. The X-axis movement is segmented into ten motion phases, including idle, warm-up, sweeps, fast and slow traverses, edge passes, stabilization, and final stop. Smartphone 1 exhibits a smoother and more uniform response, whereas Smartphone 2 captures stronger high-frequency components and orientation-dependent variations. The figure illustrates how even single-axis mechanical motion produces distinct, device-specific magnetic patterns.}
    
\end{figure}

For each synchronized time step, the combined input vector, as of equation (8), contains the three magnetic axes and one acoustic value from each smartphone, yielding an eight-dimensional observation:

\begin{equation}
\mathbf{u}_t =
\left[
x^{(1)}_t,\,
y^{(1)}_t,\,
z^{(1)}_t,\,
a^{(1)}_t,\,
x^{(2)}_t,\,
y^{(2)}_t,\,
z^{(2)}_t,\,
a^{(2)}_t
\right]
\end{equation}

The synchronized sequence is then partitioned into overlapping windows of length $L$, as shown in equation (9) and algorithm (1), where each window captures a short interval of printer activity and serves as one training instance for the learning model. The $j$-th input window is defined as

\begin{equation}
\mathbf{X}_j =
\left[
\mathbf{u}_{j},\,
\mathbf{u}_{j+1},\,
\dots,\,
\mathbf{u}_{j+L-1}
\right]
\end{equation}

\begin{algorithm}[t]
\caption{Multi-Sensor Data Synchronization}
\begin{algorithmic}[1]
\Require Two datasets \( D_1 \), \( D_2 \) with timestamps \( T_1 \), \( T_2 \)
\Ensure Synchronized dataset \( D_{\text{sync}} \) with a common time axis

\State Sort \( D_1 \) by \( T_1 \), \( D_2 \) by \( T_2 \)
\State Define unified time axis \( T_{\text{sync}} = T_1 \cup T_2 \)

\ForAll{timestamp \( t \in T_{\text{sync}} \)}
    \State Find \( t_i, t_{i+1} \in T_1 \) s.t. \( t_i \leq t \leq t_{i+1} \)
    \State Interpolate: 
    \[
    S_1(t) = S_1(t_i) + \frac{(S_1(t_{i+1}) - S_1(t_i))(t - t_i)}{t_{i+1} - t_i}
    \]
    \State Find \( t_j, t_{j+1} \in T_2 \) s.t. \( t_j \leq t \leq t_{j+1} \)
    \State Interpolate: 
    \[
    S_2(t) = S_2(t_j) + \frac{(S_2(t_{j+1}) - S_2(t_j))(t - t_j)}{t_{j+1} - t_j}
    \]
\EndFor

\State Construct \( D_{\text{sync}} \) from interpolated values
\State \Return \( D_{\text{sync}} \)
\end{algorithmic}
\end{algorithm}

This window-based representation is important because G-code execution is sequential by nature. Rather than relying on isolated sensor readings, the model learns from short temporal segments that preserve movement transitions, extrusion behavior, and modality-specific patterns over time.

Acoustic signals were recorded at 44.1 kHz (16-bit PCM), while magnetometer data were sampled at 100 Hz using the Android SensorManager API. Both smartphones recorded independently without hardware synchronization.
Because the two smartphones acquire acoustic and magnetic signals independently, the raw measurements do not share an aligned and or synchronized temporal reference. Each device collects its own timestamp sequence, sampling irregularities, amplitude scale, and sensor-specific noise characteristics. Therefore, before the collected signals can be used for learning, they must undergo temporal signal analysis to transform the raw dual-device recordings into a unified and physically consistent multimodal representation of printer activity.

We first sort the datasets according to their timestamps and project both sensing streams onto a common temporal axis. For each time point on this shared axis, missing values are estimated through interpolation so that the measurements from Smartphone~1 and Smartphone~2 correspond to the same physical interval of the printing process. This synchronization step is essential because the subsequent learning model does not operate on isolated device-specific traces; rather, it learns from the joint evolution of magnetic and acoustic emissions across time. Without temporal alignment, even small sampling offsets between the two smartphones would distort the cross-modal relationships required for reliable G-code reconstruction.

After synchronization, the signals are normalized and smoothed to suppress sensor-dependent scale variations, transient spikes, and environmental interference while preserving the dominant motion-related structure of the printer emissions. In this work, preprocessing is not treated as a simple cleaning stage, but as a mechanism for enhancing \textit{discriminative temporal features} embedded in the side-channel traces. These features include periodic motions, local transitions caused by nozzle movement, and recurrent patterns associated with extrusion and directional traversal. By suppressing irrelevant fluctuations while retaining these informative temporal characteristics, the preprocessing pipeline improves the separability of printer-induced activity patterns prior to model training.

The necessity of this step is evident in Figure 4, which presents magnetic measurements captured by the two smartphones during isolated X-axis motion. Although both devices observe the same underlying mechanical behavior, the resulting waveforms differ in amplitude, smoothness, and high-frequency variation due to sensing position, device orientation, and local channel effects. These device-dependent differences demonstrate that raw magnetic observations cannot be directly treated as a single coherent stream. Instead, they require synchronization and normalization so that the common motion-driven structure can be retained while device-specific distortions are reduced.

A similar effect is also evident for the acoustic modality. The raw traces contain transient spikes, background noise, and irregular fluctuations that partially obscure the periodic behavior generated by the nozzle motion. After smoothing and denoising, however, the waveform reveals a clearer temporal pattern corresponding to repeated printer movement during filament extrusion. This conveys how temporal signal analysis improves the observability of the underlying process dynamics and makes the acoustic modality more suitable for downstream learning.

For each synchronized time step, the preprocessed measurements are fused into a single eight-dimensional observation vector, as defined in equation (8), consisting of the three magnetic axes and one acoustic value from each smartphone. The resulting sequence is then segmented into overlapping windows of length $L$, as defined in equation (9). Each window captures a short but meaningful interval of printer activity and serves as one training instance for the learning model. This sliding-window representation preserves local temporal dependencies and cross-modal interactions, allowing the model to associate the enhanced discriminative temporal features with the corresponding G-code command patterns.

Taken together, synchronization, smoothing, normalization, and window-based segmentation convert noisy, device-dependent side-channel traces into a coherent multi-modal sequence representation. This representation forms the basis for the subsequent learning stage, where the CNN-LSTM model uses the enhanced temporal structure of the acoustic and magnetic signals to infer the executed G-code commands.

\subsection{Training Hybrid LSTM-CNN}
After temporal signal analysis, synchronization, normalization, smoothing, and sliding-window segmentation, the resulting multi-modal windows are used to train the proposed hybrid LSTM-CNN model for G-code command reconstruction. The goal of this learning stage is to map synchronized acoustic and magnetic side-channel observations to the corresponding printer command lines executed during fabrication. The CNN module consists of two 1D convolutional layers (filters = 32 and 64, kernel size = 3) followed by max-pooling. The output is fed into an LSTM layer with 128 hidden units, followed by a fully connected layer for motion classification

\begin{figure}[t]
  \centering
  \begin{minipage}[b]{0.46\columnwidth}
    \centering
    \includegraphics[width=\linewidth]{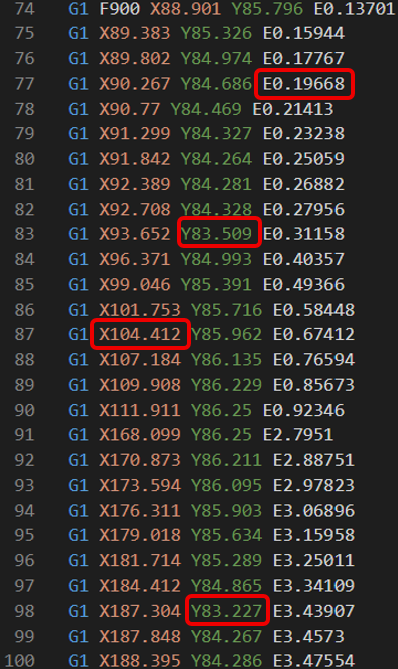}
    \par\small (a) Snippet of the Original G-code.
  \end{minipage}\hspace{0.02\columnwidth}%
  \begin{minipage}[b]{0.47\columnwidth}
    \centering
    \includegraphics[width=\linewidth]{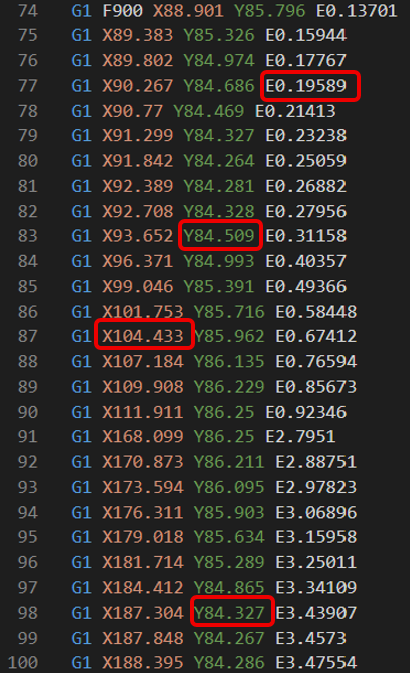}
    \par\small (b) Snippet of the Reconstructed G-code.
  \end{minipage}
  \caption{The red rectangles show the differences in the printer’s command before and after the attack
in speed, extrusion amount, and axial positions, respectively, to illustrate the accuracy of the G-code unauthorized accessibility after the attack.}
  \label{fig:pair}
\end{figure}

We also used CNN layers to detect spatial and spectral features of emissions that are produced during the printing process, whereas LSTMs were used to capture the time-dependent aspects of these emissions and how they relate to one another through the various stages of printing, such as the printer moving, idle motions, and non-extrusive periods. A model based upon this hybrid architecture has been developed to enable the model to build a relationship between the emissions and G-code, giving us the capability of reconstructing the original G-code from side-channel emissions. We constructed a window-level dataset of 24{,}000 samples, where each sample corresponds to a fixed-duration window of synchronized smartphone acoustic and magnetometer signals. For each window, we compute statistical descriptors: the mean and standard deviation of the acoustic signal (\texttt{acoustic\_mean}, \texttt{acoustic\_std}) and of the magnetometer \(x/y/z\) axes (\texttt{mag\_\*\_mean}, \texttt{mag\_\*\_std}), yielding 8 input features. Each window is labeled with a motion primitive (\(\texttt{motion\_label} \in \{\texttt{IDLE}, \texttt{UP}, \texttt{DOWN}, \texttt{LEFT}, \texttt{RIGHT}, \texttt{Z\_UP}, \texttt{Z\_DOWN}\}\)) and an extrusion indicator (\texttt{is\_extruding}). Motion increments (\texttt{dx\_mm}, \texttt{dy\_mm}, \texttt{dz\_mm}) and extrusion increments (\texttt{de\_mm}) are derived from the commanded toolpath and used both for supervision and for reconstructing a G-code-like command stream. In this dataset, \texttt{motion\_label} is deterministically derived from (\texttt{dx\_mm}, \texttt{dy\_mm}, \texttt{dz\_mm}), and \texttt{is\_extruding} is equivalent to whether \(\texttt{de\_mm} > 0\).
Additionally, the model is trained using mini-batch gradient descent with the Adam optimizer, and early stopping is employed based on validation performance to prevent overfitting. Dropout regularization is applied within both convolutional and LSTM layers to improve generalization under varying environmental noise conditions and device-specific sensing variations. The dataset is split into training and validation subsets following an 80/20 ratio, ensuring that the model is evaluated on unseen segments of printer activity. To avoid temporal data leakage, the dataset is partitioned at the print-session level rather than at the window level. This ensures that windows originating from the same print execution do not appear in both training and validation sets. As a result, the model is evaluated on entirely unseen print sessions, providing a more realistic estimate of attack performance.
After convergence, the trained hybrid model achieves a training accuracy of 97.99\% on motion classification, demonstrating its ability to reliably learn the relationship between multi-modal side-channel emissions and underlying printer commands. The high accuracy indicates that the combined acoustic and magnetic features, when processed through the CNN–LSTM architecture, provide sufficient discriminative power to distinguish between fine-grained motion primitives and extrusion states. We also draw a comparison of our model training with other available machine learning methods in Table 1.

\begin{table}[!t]
    \centering
    \caption{Model Performance Metrics}
    \renewcommand{\arraystretch}{0.95} 
    \setlength{\tabcolsep}{3pt} 
    \rowcolors{2}{gray!15}{white}
    \resizebox{0.9\columnwidth}{!}{ 
    \begin{tabular}{lccccc}
        \hline
        \rowcolor{gray!30}
        \textbf{Model} & \textbf{Acc.} & \textbf{Prec.} & \textbf{Rec.} & \textbf{F1} & \textbf{AUC} \\
        \hline
        SVM                 & 85.4  & 84.2  & 83.9  & 84.0  & 89.5  \\
        Random Forest       & 90.8  & 90.3  & 89.7  & 90.0  & 94.3  \\
        CNN                 & 94.1  & 93.8  & 93.5  & 93.6  & 96.8  \\
        LSTM                & 95.2  & 94.8  & 94.5  & 94.6  & 97.3  \\
        \textbf{CNN-LSTM (Our model)} & \textbf{97.99} & \textbf{98.5} & \textbf{98.6} & \textbf{98.5} & \textbf{99.2} \\
        \hline
    \end{tabular}}
    \label{tab:model_performance}
\end{table}

During inference, the predicted motion labels, extrusion indicators, and estimated motion increments are sequentially aggregated to reconstruct a G-code-like command stream. Specifically, consecutive window-level predictions are merged to form continuous toolpath segments, where directional commands (e.g., \texttt{G1 X Y Z}) and extrusion updates (\texttt{E}) are synthesized based on the predicted increments. This reconstruction process transforms the learned side-channel patterns into a structured representation of the printing process, enabling accurate recovery of the original G-code commands from non-invasive, multi-modal observations. It is important to note that our reconstruction focuses on motion-related components of G-code, including directional movement, extrusion state, and motion increments.

\section{Discussion}
In this section, we explain our attack results and the comparison of the accuracy level of the reconstructed G-code with the original one. Additionally, we describe the attack scenario where we changed the location of our attacker to a new environment and a new 3D printer unit within the same brand to implement our attack with our original method of attack that we discussed in Section III.

\subsection{Outcome comparison of the main and reconstructed G-code}

To evaluate the effectiveness of the reconstruction process, we compare the original G-code with the reconstructed G-code at both the command level and the geometric level, as illustrated in Figures 5 and 6, respectively.

\begin{figure}[!t]
    \centering
    \includegraphics[width=1.0\linewidth]{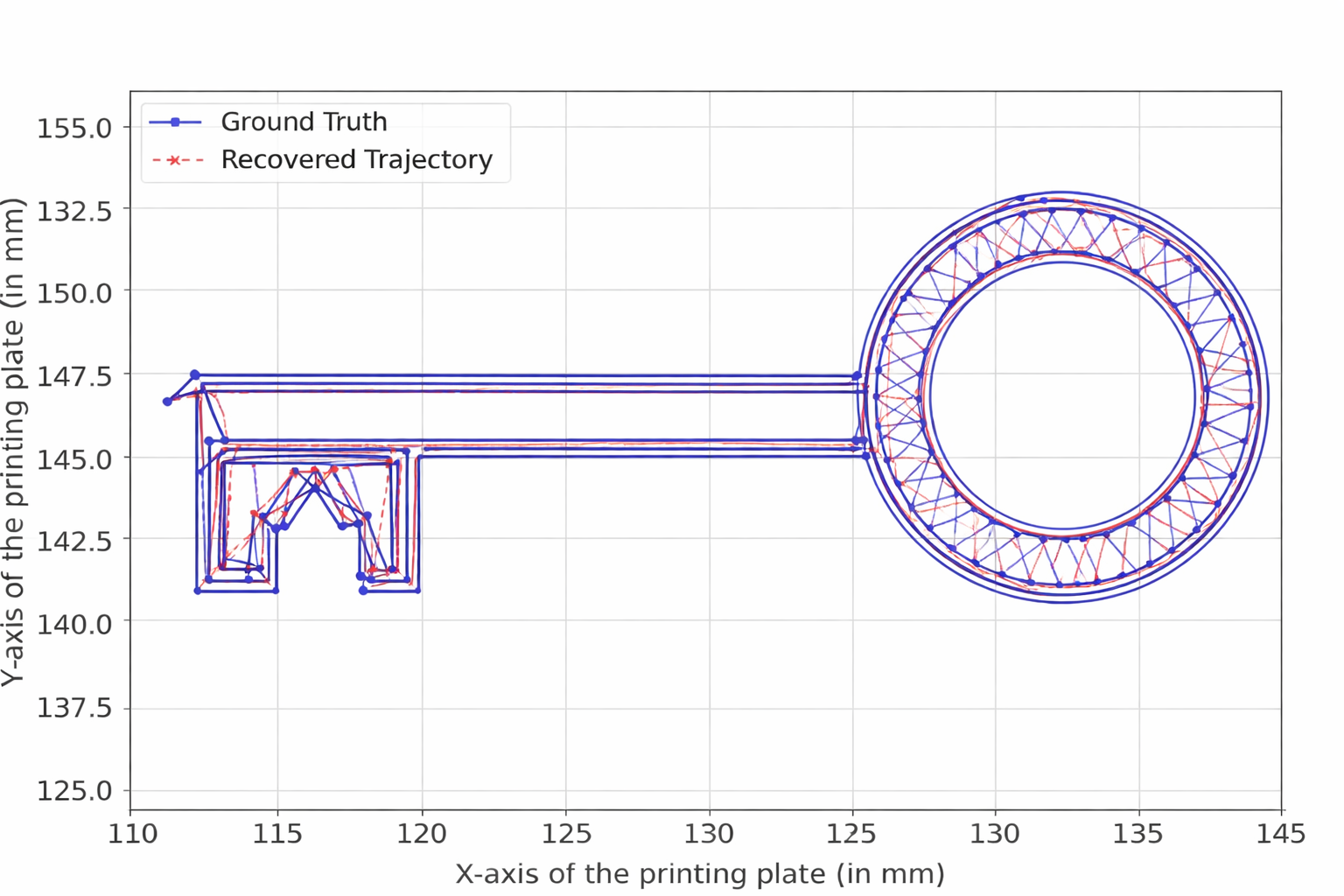}
    \caption{Comparison of the original 3D model and the reconstructed shape from G-code, demonstrating reconstruction accuracy.}
    
\end{figure}

Figure 5 presents a side-by-side comparison of selected G-code commands. The highlighted red regions indicate minor discrepancies in key parameters, including feed rate (F), extrusion value (E), and positional coordinates (X, Y). These differences are minor and do not significantly alter the overall toolpath. Such variations can be attributed to noise in the sensing process and approximation errors in the reconstruction model.

To further assess the impact of these deviations, Figure 6 compares the original printing trajectory as the ground truth with the reconstructed trajectory. The reconstructed path  follows the original geometry, preserving both linear and curved structures with high fidelity. Minor deviations are observed in complex regions, particularly in sharp transitions and circular patterns.

To quantitatively evaluate the reconstruction performance, we compute the command-level accuracy by performing a line-by-line comparison between the original and reconstructed G-code sequences, as illustrated in Figure 5.

We let $N_{\text{total}}$ denote the total number of G-code command lines in the original file, and $N_{\text{matched}}$ represent the number of correctly reconstructed command lines. A reconstructed command is considered correct if its key parameters, including motion direction and numerical values, such as $X$, $Y$, $Z$, $E$, and $F$, match those of the original command within a predefined tolerance threshold $\epsilon$.

Formally, the reconstruction accuracy is defined as:

\begin{equation}
\text{Accuracy (\%)} = \frac{N_{\text{matched}}}{N_{\text{total}}} \times 100
\end{equation}

To account for minor deviations caused by noise in side-channel measurements and approximation errors in the learning model, numerical parameters are compared using a tolerance-based criterion. Specifically, a reconstructed command $i$ is considered a match if:

\begin{equation}
\left| p_i^{\text{orig}} - p_i^{\text{recon}} \right| \leq \epsilon, \quad \forall p \in \{X, Y, Z, E, F\}
\end{equation}

where $p_i^{\text{orig}}$ and $p_i^{\text{recon}}$ denote the original and reconstructed parameter values for the $i$-th command, respectively.

Using this metric, our approach achieves a reconstruction accuracy of \textbf{98.89\%}, indicating that the vast majority of G-code commands are successfully recovered. As shown in Figure 5, the remaining discrepancies are limited to small variations in parameters such as extrusion and feed rate, which do not significantly affect the overall toolpath geometry.

These results demonstrate that the proposed reconstruction approach is capable of recovering G-code with high precision. Despite slight numerical differences at the command level, the reconstructed output maintains strong geometric consistency with the original model, validating the effectiveness of the method for practical applications.

\subsection{Data Collection Integrity in Multi-Printer Environments} 

To ensure the accuracy and source-directionality of the observed data, we implemented an experiment with two printers, as shown in Figure 7. The left-side printer, LulzBot TAZ \cite{sansing2014taz}, was the target attack 3D printer, and a second 3D printer, Creality \cite{Creality}, was nearby as a simulated interference printer. The main reason we chose the Creality 3D printer to work simultaneously with our targeted 3D printing unit was due to its popularity, availability in commercial use, and low costs \cite{sahat2023study}.

\noindent
\begin{figure}[!t]
    \centering
    \includegraphics[width=1.0\linewidth]{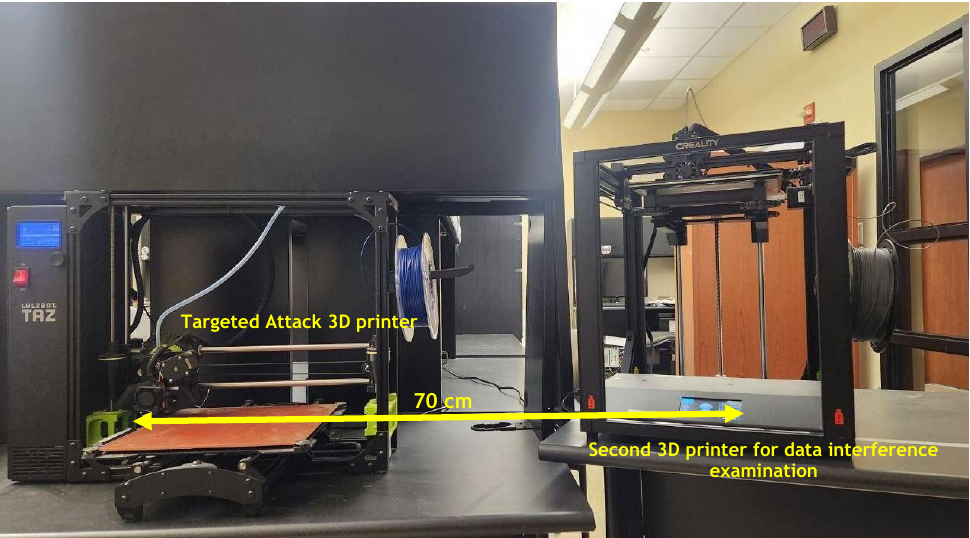}
    \caption{Two 3D printers working at the same time for data interference examination} 
    
\end{figure}

This setup enabled us to evaluate whether other printers that operate simultaneously as adjacently positioned devices interfere with the data collection process.

Based on our observations, the LulzBot TAZ \cite{sansing2014taz} printer has larger motors, stiffer mechanical components, and larger cooling fans that emit noise at a louder rating than a more compact Creality printer \cite{Creality}. In short, the amplitude of sound from the targeted printer was much stronger. Given that acoustic intensity decreases with the square of the distance $I \propto \frac{1}{d^2}$ and smartphones use directional microphone sensitivity, both geometric and hardware limitations inherently reduce the amount of sound energy radiating from the Creality printer. In addition to these limitations, physical obstructions (e.g., aluminum gantries and housing panels) associated with the Creality printer reduced the sound even further before reaching the sensor. As shown in Figure 9, the targeted attack 3D printer, Lulzbot, exhibited stronger signal spikes and was clearly distinguishable from the other 3D printer operating at the same time. Compared to the 3D printer targeted for attack, the Creality printer's signals were significantly weaker due to the distance. 

In our setup, the LulzBot printer's motors were located at a specific angle from the smartphone's magnetometer. As a result, the magnetic signals captured from the LulzBot printer were strong and consistent. In contrast, the Creality printer’s motors were positioned more randomly without a calculated degree angle and partially blocked by the printer’s own metal frame. This structure likely absorbed or distorted the already weaker magnetic fields emitted by the Creality printer, making their contribution to the recorded data negligible.  

\subsection{Attack setup in a new environment and a new
3D printing unit}

To evaluate generalization beyond the original controlled scenario, we executed the attack using a different unit of the same 3D printer model \cite{taz6} in a different physical environment.

\begin{table}[t]
\centering
\scriptsize
\caption{Reconstructed objects via the 2nd 3D printer in a different environment and background noise.}
\label{tab:reconstructed_second_printer}

\setlength{\tabcolsep}{2pt}
\renewcommand{\arraystretch}{0.9}

\begin{tabular}{
    >{\raggedright\arraybackslash}m{0.22\linewidth}
    >{\raggedright\arraybackslash}m{0.43\linewidth}
    >{\centering\arraybackslash}m{0.25\linewidth}
}
\hline
\textbf{Model} & \textbf{Metrics} & \textbf{Reconstructed objects} \\
\hline

Gear Icon &
Print Duration: 46 mins; Steps: 29,369; Accuracy: 98.05\%; Complexity: High
&
\includegraphics[width=0.75\linewidth]{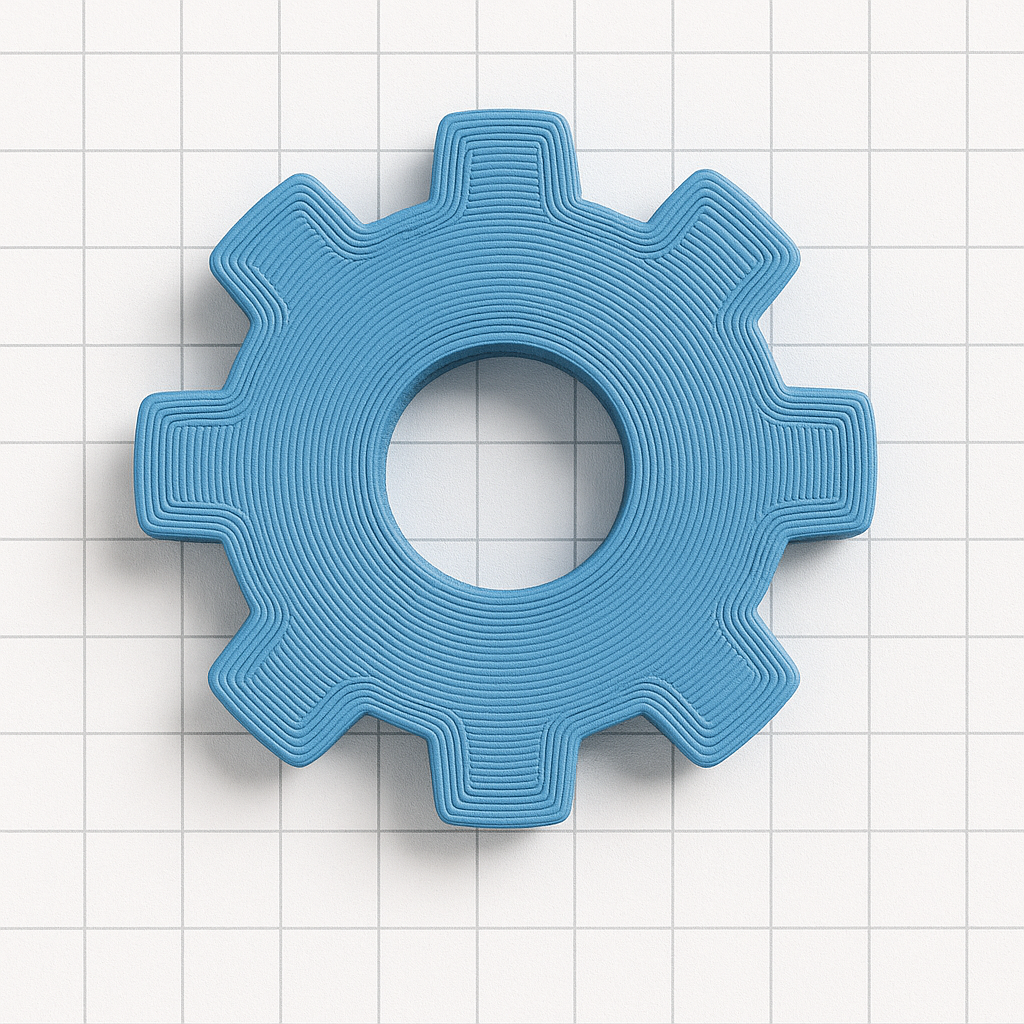}
\\
\hline

Turbine blades &
Print Duration: 68 min; Steps: 30,412; Accuracy: 98.10\%; Complexity: High
&
\includegraphics[width=0.75\linewidth]{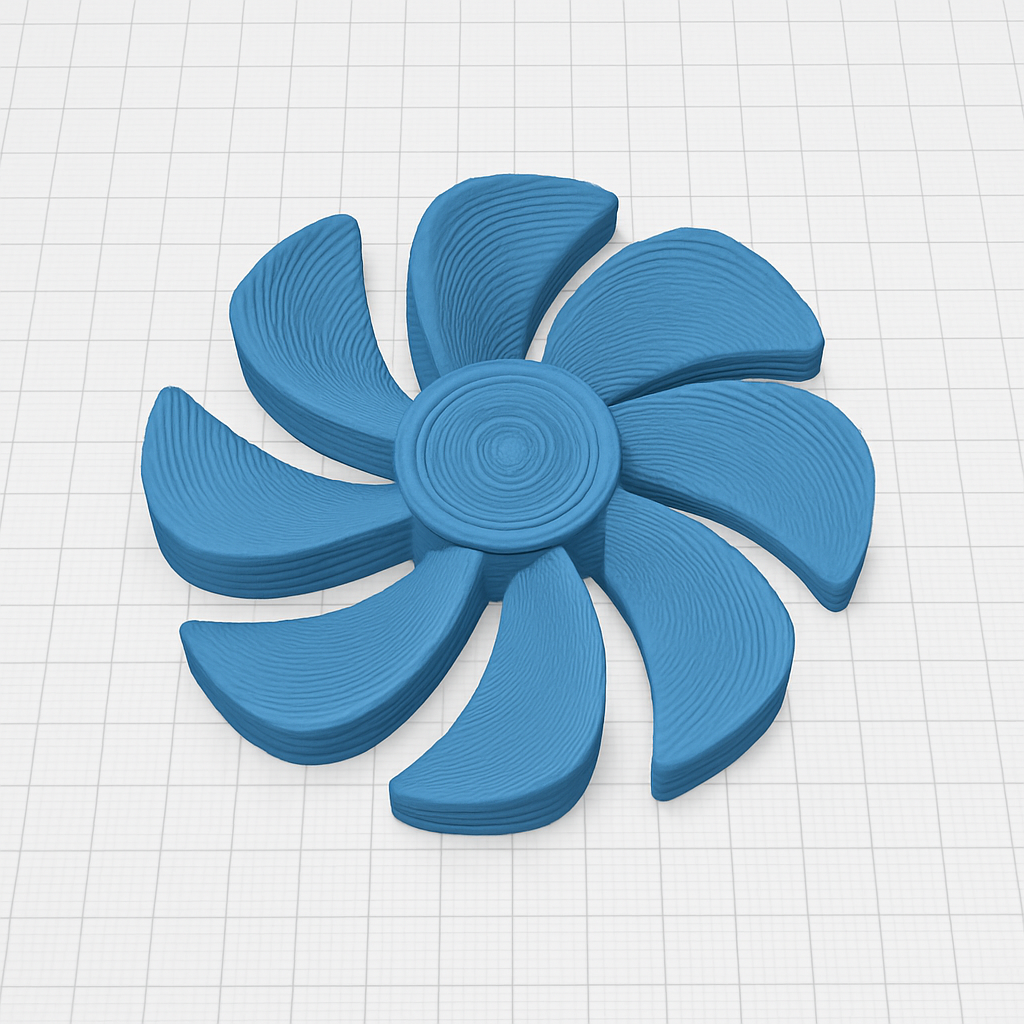}
\\
\hline

Wrench &
Print Duration: 39 min; Steps: 27,105; Accuracy: 98.35\%; Complexity: High
&
\includegraphics[width=0.75\linewidth]{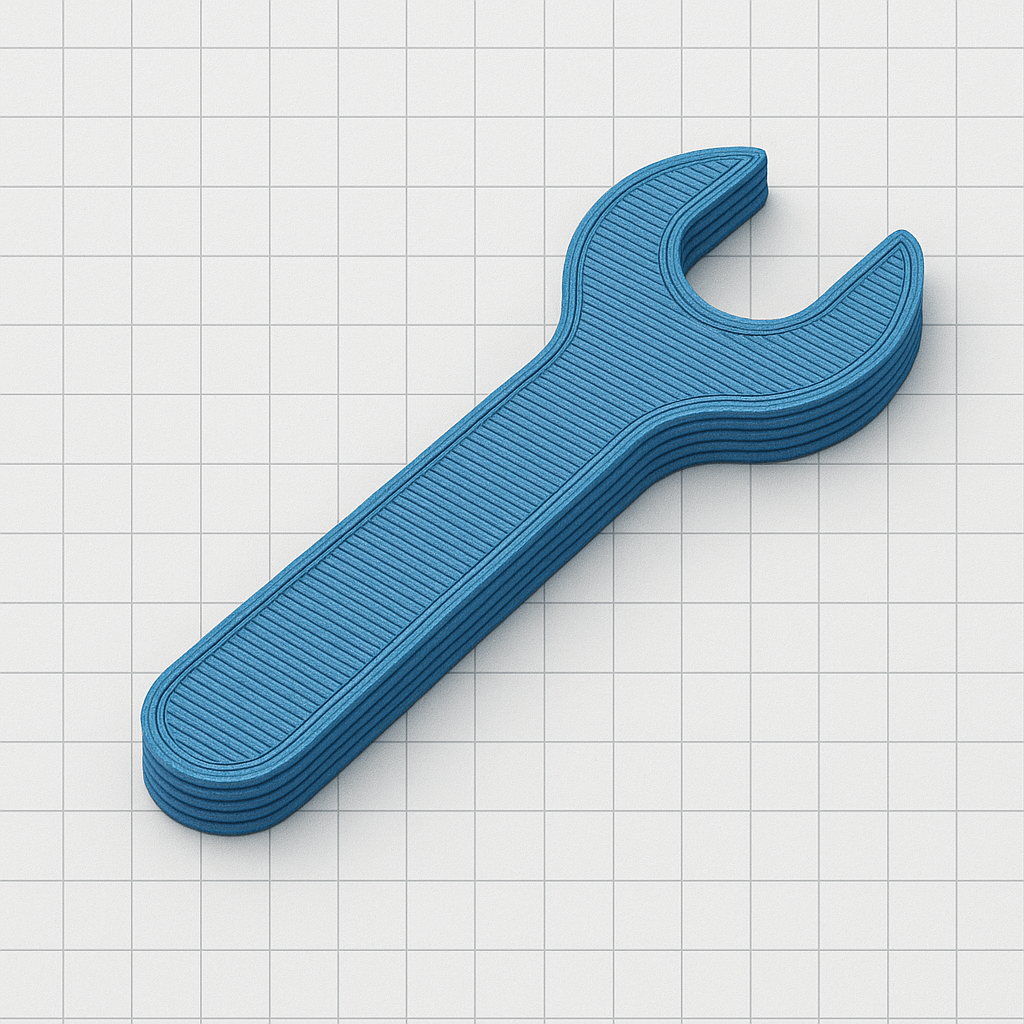}
\\
\hline

\end{tabular}
\end{table}

The smartphones were positioned in a non-line-of-sight configuration by replicating the original setup at distances of 70 cm and 65 cm from the printer's surface bed. For the second attack scenario, we followed the same data-collection method, pre-processing pipeline, and parameter-variation strategy used in the original attack described in Section IV. The positioning and orientation of the sensor are illustrated in Figure 8 and the relevancy of the distance and angle of the phones are depicted in Figure 10. The only difference in the new attack was to further increase the distance between the smartphones and the 3D printing unit. 

\begin{table*}[t]
\centering
\caption{Reconstruction accuracy at different sensor distances. 
Each row represents the number of independent print sessions ($n=12$), 
mean reconstruction accuracy, sample standard deviation (SD), 
95\% confidence interval (CI), mean acoustic SNR, magnetic-field magnitude, 
and the resulting signal-quality classification.}
\label{tab:reconstruction-distance}

\begin{tabular}{@{}cccccccc@{}}
\toprule
Distance (cm) & $n$ & Mean Acc. (\%) & SD (\%) & 95\% CI (\%) & Acoustic SNR (dB) & $|\mathbf{B}|$ (µT) & Signal Quality \\ 
\midrule
30  & 12 & 99.2 & 0.4 & [98.9, 99.5] & 26.4 & 47.8 & Strong \\
45  & 12 & 98.9 & 0.6 & [98.5, 99.3] & 23.1 & 44.2 & Strong \\
\textbf{60} & \textbf{12} & \textbf{98.4} & \textbf{0.7} & \textbf{[97.9, 98.9]} & \textbf{19.6} & \textbf{40.5} & \textbf{Strong} \\
\textbf{70}  & \textbf{12} & \textbf{98.0} & \textbf{0.9} & \textbf{[97.3, 98.6]} & \textbf{17.8} & \textbf{37.9} & \textbf{Strong} \\
75  & 12 & 92.4 & 1.8 & [91.3, 93.6] & 11.5 & 31.2 & Medium \\
100 & 12 & 78.8 & 2.6 & [76.9, 80.7] & 5.2  & 24.7 & Weak \\
130 & 12 & 69.2 & 3.1 & [67.0, 71.5] & 1.6  & 19.3 & Very Weak \\
$>$150 & 12 & $<$65.0 & 3.8 & [60.7, 64.9] & $<0$  & 16.4 & Unreliable \\
\bottomrule
\end{tabular}

\vspace{1ex}
\footnotesize{
Signal-quality thresholds: Strong ($\text{SNR} \ge 20$ dB),
Medium ($10 \le \text{SNR} < 20$ dB),
Weak ($0 \le \text{SNR} < 10$ dB),
Very Weak ($\text{SNR} < 0$ dB).
}
\end{table*}

We aimed at reconstructing the same key shape as illustrated in Figure 6. We draw up an accuracy comparison in Table 3. As shown, accuracy stays at 98-99\% within a short to moderate range, while the acoustic SNR is still strong. Beyond 100 cm, both SNR and magnetic-field strength sharply drop, so accuracy falls into the very weak range. Distance greater than 150 cm, signal quality becomes unreliable, and reconstruction accuracy drops below 65\%. Each distance setting is evaluated over $n = 12$ independent print sessions. Each session corresponds to a full print execution of the same object under identical printer settings but with independently recorded sensor data. This setup allows us to isolate the impact of distance on signal quality and reconstruction performance. Additionally, we implemented our attack on different objects, such as a gear, turbine, and a wrench, and the details of our reconstructed G-code and accuracy level are illustrated in Table 2. 

While the proposed approach demonstrates high reconstruction accuracy under the evaluated conditions, limitations remain. First, performance degrades significantly at longer distances where the signal-to-noise ratio becomes low. Second, the multi-printer interference evaluation is limited to scenarios where the target printer produces dominant signals. Addressing these limitations is an important direction for future works.

\noindent
\begin{figure}
    \centering
    \includegraphics[width=1.0\linewidth]{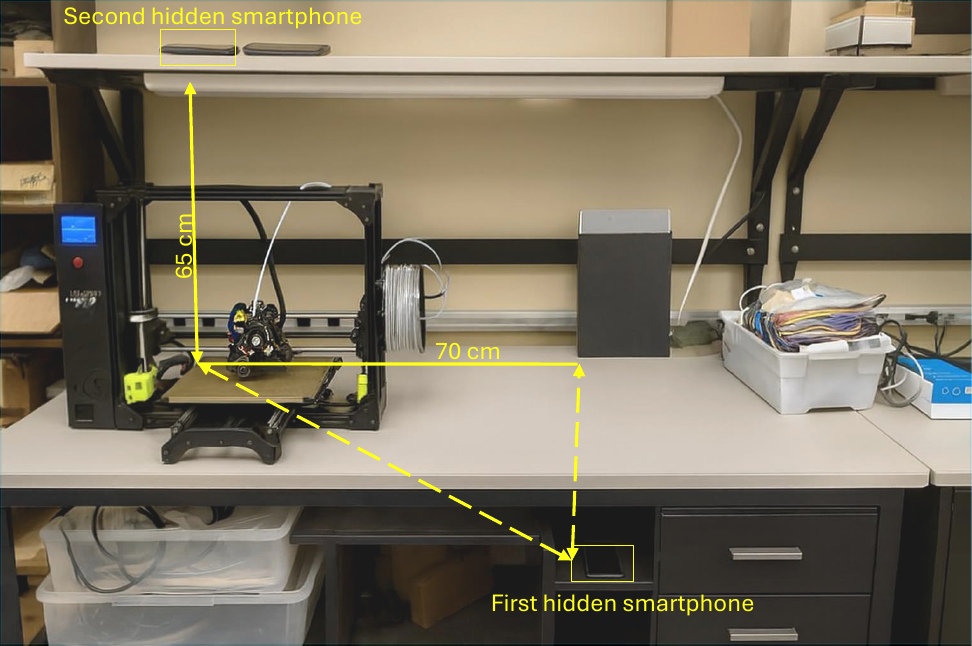}
    \caption{Attack on the 3D printer in a new environment with a different 3D printer.} 
    \label{fig:yourlabel}
\end{figure}

\begin{figure*}[t]
    \centering
    \includegraphics[width=\textwidth]{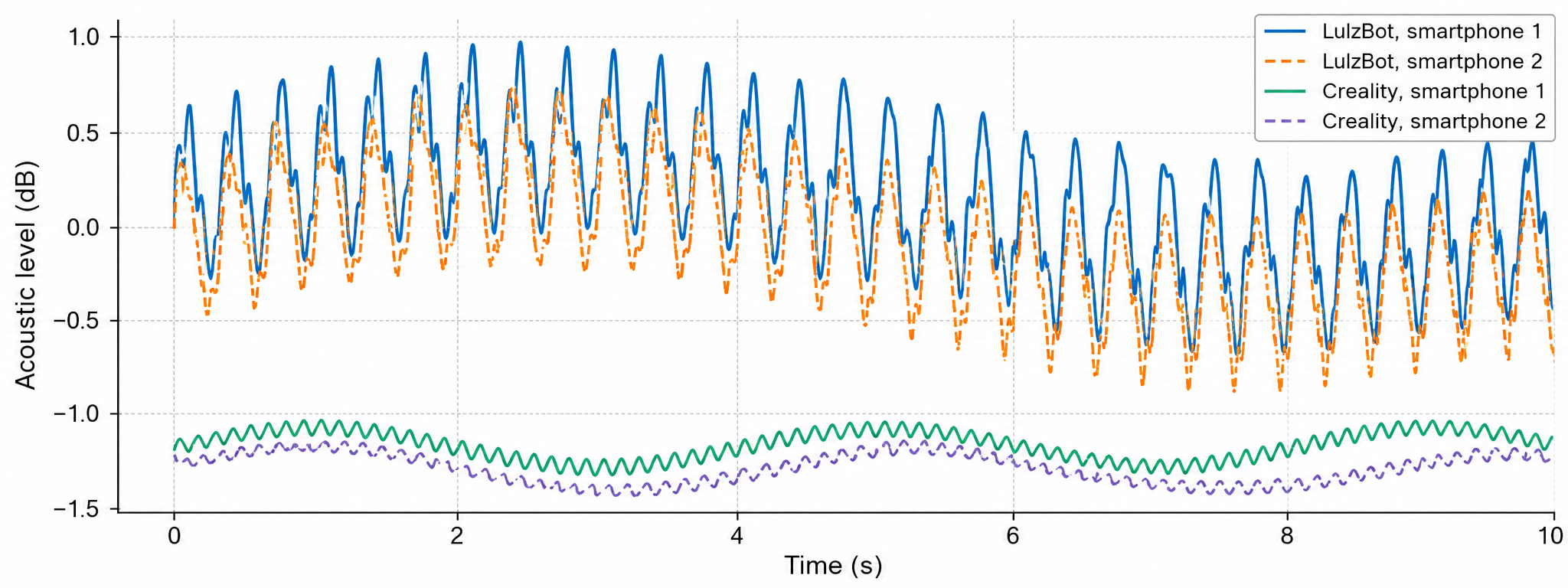}
    \caption{Acoustic emission signals captured during 3D printing (only moving on X-axis) with two different types of machines (LulzBot vs Creality). The LulzBot printer displays much larger amplitude acoustic features, including the existence of clear periodic peaks associated with motor surges, mechanical resonance, and tension events. The Creality printer had overall much smaller amplitude signals, with limited patterns. The large contrast in signal magnitude enabled us to straightforwardly isolate and remove the Creality trace during post-processing, allowing us to focus on the dominant LulzBot signal.}
    \label{fig:acoustic_merged}
\end{figure*}

\section{Countermeasures} 
Following our successful attack on the 3D printer via side channels, we identified that multiple layers of countermeasures are needed to reduce side channel attack risk on 3D printers and safeguard G-code confidentiality. We believe that these adjustments and solutions will reduce the risk of unauthorized data extraction by addressing vulnerabilities at the hardware, software, and operational levels.

\subsection{Acoustic and Magnetic Emission Shielding}

Reducing the emission of exploitable signals is among the most successful strategies to stop side-channel attacks. The following strategies mainly implement the blocking methods of the attacks:

\begin{enumerate}
    \item  Encapsulating the 3D printer in a soundproof enclosure with the implementation of materials that absorb or scatter sound waves; this will block the spread of acoustic sounds. Similarly, including regulated random background noise also prevents the scattering of unintended signals.
    \item  For the purpose of blocking magnetic data to be collected by outside devices, Faraday cages \cite{pula2019analysis} or conductive enclosures around the printer help reduce electromagnetic signal leakage, making it more difficult for attackers to follow the trend of magnetic emissions. 
\end{enumerate}

\subsection{Randomization of Printing Parameters}

Randomizing printing parameters will limit an attacker's capacity to link obtained signals with particular printer actions. In other words, it is an example of a hindrance to the straight path.  However, it will add to the printing time and complexity of the G-code commands as there are additional printing parameters, which will help the original G-code to remain safe or prevent unintended magnetic and acoustic emissions. The following methods can execute this implementation:

\begin{enumerate}
    \item Extrusion Path Obfuscation: we can block an attacker's malicious actions by injecting slightly different toolpath sequences or layer deposition orders into the G-code without compromising print quality; this action will throw off signal consistency, which misleads possible attackers' data collection, so that they can never reach the original G-code by analyzing counterfeit data. 
    \item Randomizing tiny motor movement variations, acceleration profiles, and print speeds makes it more difficult for attackers to extract precise G-code commands from recorded emissions.  
\end{enumerate}

\subsection{Active Jamming and Anomaly Detection}

The use of active jamming and anomaly detection is another solution to prevent attackers from following the trend of consistent unintended data emission \cite{ahsan2023sok, beckwith2021needle, yang2022online}. This method can be implemented using the following approach:

The deployment of active noise emitters that produce white noise across both the acoustic and electromagnetic spectra can interfere with attackers' signal collection. This method interferes with the original data emitted unintentionally by the 3D printer. Even if the attacker follows these data and attempts to collect it remotely, he will ultimately find random noise levels gathered that have no relation to the printed final object. 

\begin{figure}[!t]
    \centering
    \includegraphics[width=1.0\linewidth]{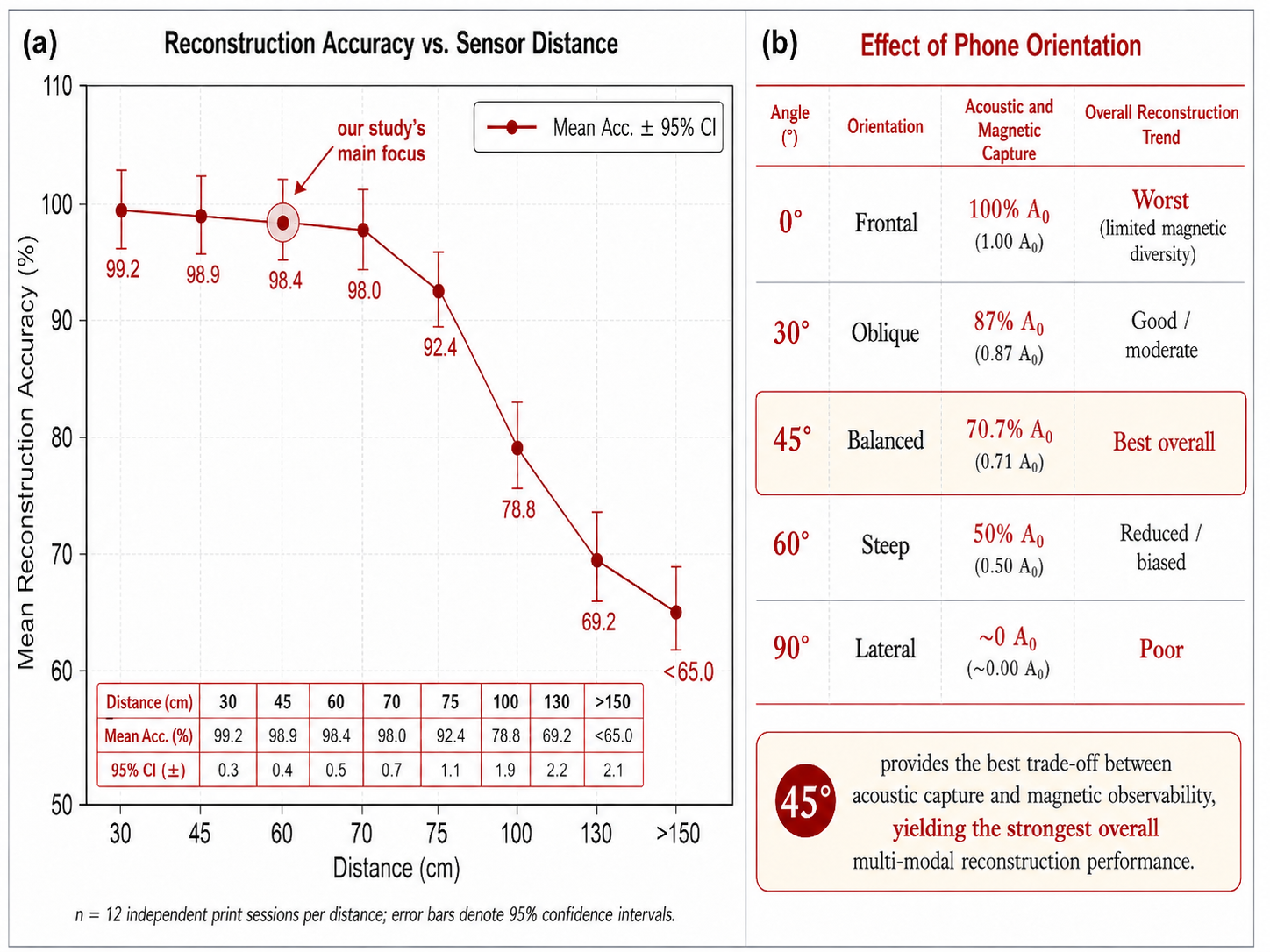}
    \caption{(a) Mean reconstruction accuracy decreases as the sensor distance from the printer increases, with the proposed setup focusing on the 60 cm non-contact sensing distance. Error bars indicate 95\% confidence intervals across 12 independent print sessions. (b) Effect of phone orientation on acoustic and magnetic signal capture, showing that a 45° placement provides the best trade-off between acoustic capture and magnetic observability, resulting in the strongest overall multi-modal reconstruction performance.} 
    
\end{figure}

\section{Related work}
Several studies on cyber-physical attacks have
highlighted the vulnerability of attackers gaining unauthorized access to 3D printer G-code commands. In this section, we discuss what works have already
been studied.
Prior work has shown that magnetic and acoustic emissions from FDM printers leak motion information strongly enough to support reconstruction of toolpath primitives and, in some cases, G-code \cite{song2016my,gatlin2021encryption,al2016acoustic,yampolskiy2016using,asgar2026quietprint,liang2022hiding}.

In comparison to prior both acoustic and magnetic pipelines that rely on close-range, carefully positioned recording and still face direction ambiguity during complex motion, our work focuses on dual-smartphone, multi-modal attack designed explicitly for realistic environments rather than merely demonstrating feasibility under constrained setups. We develop and evaluate a concealed attack using two smartphones and a dynamic noise modeling framework that incorporates environmental noise, enabling robust G-code reconstruction even in noisy settings, including scenarios where another 3D printer operates concurrently. Unlike prior smartphone-based studies that emphasize distance-driven degradation and often treat multi-device sensing as future work, we operationalize a multi-phone approach and evaluate it under non-line-of-sight placements without physical contact to the victim printer.

\section{Conclusion} 

This study investigated the security exploitability of AM systems, specifically 3D printers. We analyzed the feasibility of gaining unauthorized access to the 3D printer's G-code commands. We demonstrated how attackers can capture side-channel data by positioning smartphones in a non-line-of-sight at a specific distance and angle relative to the 3D printer, allowing them to gain unauthorized access to the final printed object's design file. As smartphones are equipped with magnetic and sound sensors, the data collection process for the attack becomes increasingly simple and inconspicuous, putting AMs at risk of manipulation. Our attack achieved a reconstruction accuracy of 98.89\% for the G-code of the key-shaped object.
Our G-code reconstruction highlights the need for awareness and preventive actions against side-channel attacks in the AM industry.

%
\IEEEpeerreviewmaketitle




%

\bibliographystyle{IEEEtran}
\bibliography{sample-base}  

\end{document}